\documentstyle[ibvs,psfig]{article}     

\def\ros{{\sl ROSAT }}
\def\etal{{\it et\,al. }}
\def\amin{\ifmmode ^{\prime}\else$^{\prime}$\fi}
\def\asec{\ifmmode ^{\prime\prime}\else$^{\prime\prime}$\fi}
\def\farcs{\hbox{$.\!\!^{\prime\prime}$}}  
\def\rxj{RX\,J0527.8--6954}

\begin{document}
 
\IBVShead{4409}{?? December 1996}

\title{HV 2554 and the supersoft X-ray source RX J0527.8--6954\footnote{
          Partly based on observations with the ESO 2.2\,m
          telescope at La Silla/Chile (MPI time).}}

\begintext

\begin{figure}
      \centering{
      \hspace*{.01cm}
      \vbox{\psfig{figure=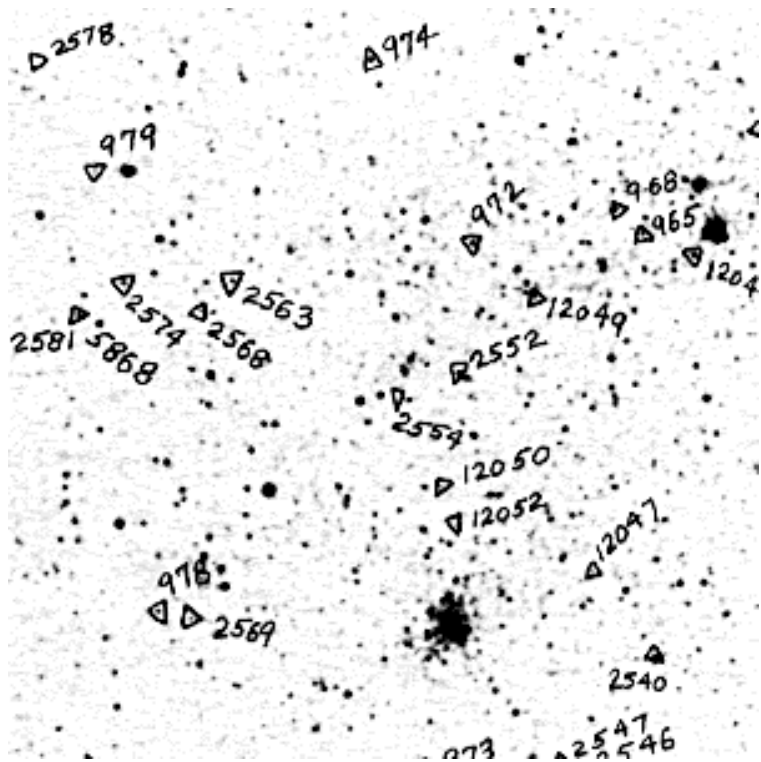,width=10.cm}}}
      \caption[hodge]{A 13\farcm5*13\farcm5 area around HV 2554 (center) 
         reproduced from the Hodge \& Wright (1967) atlas of the LMC. 
         The variable is located inside the triangle above the ``2554" mark.
         North is at the top and East to the left.}
         \label{hodge}
\end{figure}

The discovery of the supersoft X-ray source RX\,J0527.8--6954 during the
\ros first light observation (Tr\"{u}mper \etal 1991) of the Large Magellanic 
Cloud (LMC) in June 1990 has directed some attention to the optical variable
HV 2554 because its location is within the X-ray error circle of 
\rxj\, (Tr\"umper \etal 1991, Greiner \etal 1991). Later ROSAT observations
improved the X-ray position resulting in a larger offset to HV 2554
(Cowley \etal 1993, Greiner \etal 1996a,b). However, it also became clear
that there are no exact coordinates available (no SIMBAD entry) for HV 2554.
To our knowledge the only finding chart available for HV 2554 is the
Large Magellanic Cloud atlas by Hodge \& Wright (1967) (see Fig. \ref{hodge}),
but unfortunately the scale is too poor and the variable itself invisible.
While only a summary of the variability of HV 2554 is published in form of
table entries in Shapley \& Mohr (1940) (based on the investigation of only 
12 plates) and Shapley \& McKibben Nail (1955), the detailed notes of the 
Gaposchkins (C.H. Payne-Gaposchkin and S. Gaposchkin) on the brightness 
estimates of HV 2554 on 380 plates (taken between 1896 and 1954) of mainly 
the A series are unpublished.

\begin{figure}
      \centering{
      \hspace*{.01cm}
      \vbox{\psfig{figure=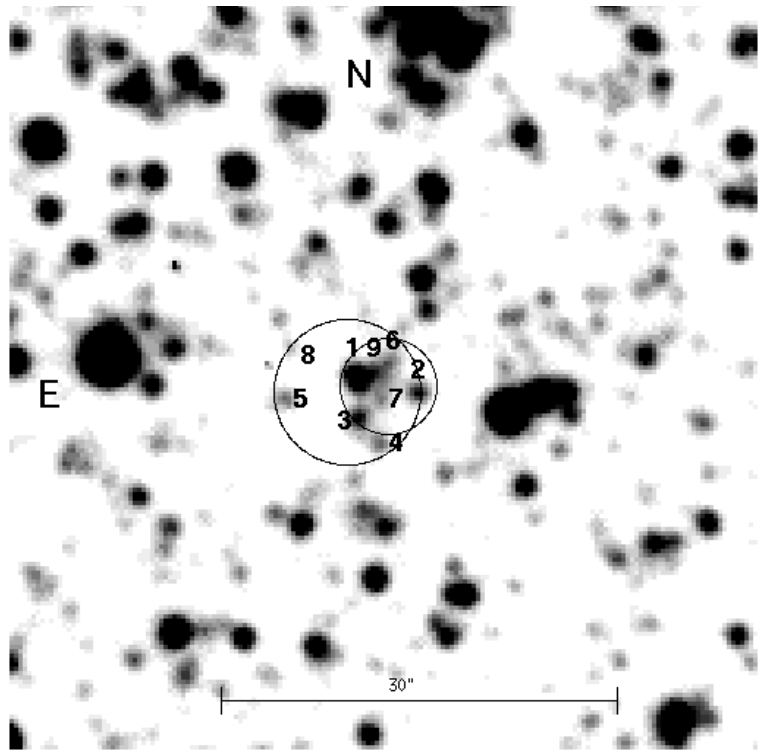,width=10.cm}}}
      \caption[ccd]{The 5\asec\, X-ray error radius (large circle) of
           RX J0527.8--6954 overplotted on a 10 min B image taken on 
           March 25, 1995 with the ESO 2.2\,m telescope at La Silla/Chile.
           The small circle denotes the best-fit astrometric position of 
           HV 2554 as determined from plate A 14531 of the Harvard plate 
           collection. Numbers denote all resolved objects within the
           X-ray error circle (large circle, Greiner \etal 1996a).}
         \label{ccd}
\end{figure}

Given these facts we went back to the original plates and
re-identified HV 2554. From the unpublished individual brightness 
estimates (recently archived by D.L. Welch and 
electronically available on http:/$\!$/www.physics.mcmaster.ca/HCO/) 
we selected four plates: two with 
HV 2554 being brightest and two plates with it being in a faint state. 
A comparison of the brightest/faintest plate pairs quickly revealed a 
clearly variable object with an amplitude consistent
with the value of $\Delta$m$\sim$1.6 mag (Shapley \& McKibben Nail 1955).
Our independent relative brightness estimates on nearly 30 further plates
are in good agreement to those of the Gaposchkins and thus confirm the 
correctness of our re-identification. The astrometry on plates showing 
HV 2554 in the bright state is overplotted on a CCD frame taken in March 1995
(small circle on Fig. \ref{ccd}) and demonstrates that its position is
within the 5\asec\, X-ray error circle of \rxj.

These findings provided the motivation to determine the pattern of optical 
variability of HV 2554 over the last six years during which \rxj\, was found to 
gradually decline in X-ray intensity (Greiner \etal 1996a,b). For this purpose,
we investigated about 140 blue plates out of the 447 plates (210 blue, 230 red)
taken between October 1990 and January 1995 within the EROS project for the
search of microlensing events of the LMC (Aubourg \etal 1993). Two different
emulsions were used in the blue passband (with filter GG385): 
IIaO during 1990--1993, and the emulsion IIIaF during Oct. 1993--1995.
While plates of both emulsion types are generally more sensitive than the
Harvard plates, the IIIaF emulsion even provides a spatial resolution 
below 2\asec, thus reaching in best cases a quality comparable to the 
CCD image shown in Fig. \ref{ccd} (seeing of 0\farcs9). As a consequence,
in most cases several or even all of the at least 6 objects within the
astrometric error circle of HV 2554 are resolved and detectable on these
EROS project plates. In addition to these plates, we have investigated single 
plates taken for other purposes in the years 1975, 1977, 1978, 1987 and 1989.
The surprising result of our analysis of all the investigated plates
was the fact that we did not find any variable object within or around the 
astrometric position of HV 2554.

The non-variability of any of these objects on the EROS Schmidt plates as
opposed to the apparent variability on the Harvard plates can be due to 
several reasons:
\begin{enumerate}
\vspace{-0.25cm}\item {\it The re-identification of HV 2554 is wrong} 
while the original measurements are of a different object. We have carefully 
checked this possibility, but can definitely exclude it. There is no other
star of the given brightness around the position marked on the 
Hodge \& Wright 
(1967) atleas, and in addition the variability pattern found on the plates
coincides with that of the unpublished notes of the Gaposchkins.
\vspace{-0.25cm}\item {\it HV 2554 has ceased to be variable} in the two 
decades between the last Harvard plates (1954) and the first EROS project 
plates (1990) (with the few other, individual plates it would be even before 
1977). Though this would be a rare circumstance, it cannot be excluded.
\vspace{-0.25cm}\item {\it HV 2554 is not intrinsically variable on the Harvard
plates.} Instead, the combination of variable seeing and different limiting 
magnitudes of the plates result in a different size of the image of the
several overlapping objects and thus counterfeits a variability. This reasoning
implies a clear prediction, namely that HV 2554 appears bright on plates
with better than average seeing and sensitivity, so that objects 2, 3 and 6 
(and probably also 4) contribute to the size of the merged image while on 
plates with bad seeing and  sensitivity only object 1 is imaged, thus resulting
in a considerably smaller size on the plate. A re-investigation of the Harvard
plates has indeed confirmed this relation between the brightness of HV 2554
and the plate quality.
\end{enumerate}\vspace{-0.25cm}

We therefore conclude that though variations are seen at first glance
on the Harvard plates, a careful look including a consideration of the effects
of different seeing, different fog level and limiting magnitude shows that 
variations of HV 2554 are marginal at best. A hint of support comes from the 
fact that the measurements on the unpublished notes from the Gaposchkins
were crossed out which usually means that they did not consider the object to be
variable in the end.
We would like to mention, however, that it is not possible to 
exclude definitely intrinsic optical variability of HV 2554.

Given the large amplitude of the X-ray decline of \rxj\, over the last
six years (a factor of 50), one is inclined to expect a correlated (either
positive or negative) variability of its optical emission. The lack of any 
obvious optical variability of objects 1 through 9 in Fig. \ref{ccd}
(though somewhat uncertain for the faint objects 6 through 9) suggests
that none of these is the optical counterpart of \rxj. Sensitive optical
observations (imaging and spectroscopy) at sub-arcsecond resolution are 
certainly required to identify \rxj.

\vspace{0.5cm} \noindent{\it Acknowledgements}:
JG is indebted to J. Guibert and E. Lesquoy for kind hospitality at the Paris 
Observatory. We are grateful to D. Welch for detailed information on the 
unpublished Gaposchkin material. JG is supported by the Deutsche Agentur 
f\"ur Raumfahrtangelegenheiten (DARA) GmbH under contract FKZ 50 OR 9201.

\references

Aubourg E., Bareyre P., Brehin S., Gros M., Lachieze-Rey M., Laurent B., 
    Lesquoy E., Magneville C., Milsztajn A., Moscoso L., Queinnec F., Rich J.,
    Spiro M., Vigroux L., Zylberach S., Ansari R., Cavalier F., Moniez M.,
   Beaulieu J.-P., Ferlet R., Grison Ph., Vidal-Madjar, Guibert J., Moreau O.,
    Tajahmady F., Maurice E., Prevot L., Gry C., 1993, Nat. 365, 623

Cowley A.P., Schmidtke P.C., Hutchings J.B., Crampton D.,
      McGrath T.K., 1993, ApJ 418, L63

Greiner J., Hasinger G., Kahabka P. 1991, A\&A 246, L17

Greiner J., Schwarz R., Hasinger G., Orio M., 1996a, A\&A 312, 88

Greiner J., Schwarz R., Hasinger G., Orio M., 1996b, in Supersoft X-ray 
Sources, ed. J. Greiner, LNP 472, p. 145

Hodge P.W., Wright F.W., 1967, Atlas of "The Large Magellanic Cloud",
   Smithonian Press

Shapley H., Mohr J., 1940, Ann. of Harvard Coll. Obs. 90, No. 1

Shapley H., McKibben Nail V., 1955, Proc. of Nat. Acad. of Sci. 41, No. 4, 185

Tr\"{u}mper J., Hasinger G., Aschenbach B., Br\"{a}uninger H.,
              Briel U.G., Burkert W., Fink H., Pfeffermann E., Pietsch W.,
              Predehl P., Schmitt J.H.M.M., Voges W., Zimmermann U.,
              Beuermann K., 
              1991, Nat 349, 579

\vspace*{10mm}

\beginauthors
JOCHEN GREINER
MPI f\"{u}r extraterrestrische Physik
Giessenbachstr. 1
D-85740 Garching, Germany
e-mail: jcg@mpe-garching.mpg.de

\vspace{-.15cm}
\authorsrightcol
MARTHA L. HAZEN
Harvard College Observatory
60 Garden Street
Cambridge, MA 02138
e-mail: mhazen@cfa.harvard.edu
\endauthors
 
\end{document}